\documentclass{ifacconf}

\usepackage{graphicx}      
\usepackage{natbib}        
\usepackage{graphics} 
\usepackage{epsfig} 
\usepackage{mathptmx} 
\usepackage{amsmath} 
\usepackage{amssymb}  
\usepackage{tikz}
\usetikzlibrary{decorations.pathreplacing}
\usepackage{pgfplots}
\usepackage{xcolor}
\usepackage{float}

\usepackage{enumitem}
\usepackage{booktabs}
\usepackage[short]{optidef}
\usepackage{subcaption}

\newtheorem{remark}{Remark}

\begin{document}
\begin{frontmatter}

\title{An Approximate Dynamic Programming Approach for Dual Stochastic Model Predictive Control } 


\author{Elena Arcari, } 
\author{Lukas Hewing, } 
\author{Melanie N. Zeilinger}

\address{Institute for Dynamic Systems and Control, ETH Z\"{u}rich, Z\"{u}rich, Switzerland,  (e-mail: earcari,lhewing,mzeilinger.ethz.ch).}

\begin{abstract}                
Dual control explicitly addresses the problem of trading off active exploration and exploitation in the optimal control of partially unknown systems.
While the problem can be cast in the framework of stochastic dynamic programming, exact solutions are only tractable for discrete state and action spaces of very small dimension due to a series of nested minimization and expectation operations. We propose an approximate dual control method for systems with continuous state and input domain based on a rollout dynamic programming approach, splitting the control horizon into a dual and an exploitation part. The dual part is approximated using a scenario tree generated by sampling the process noise and the unknown system parameters, for which the underlying distribution is updated via Bayesian estimation along the horizon. In the exploitation part, we fix the resulting parameter estimate of each scenario branch and compute an open-loop control sequence for the remainder of the horizon. The key benefit of the proposed sampling-based approximation is that it enables the formulation as one optimization problem that computes a collection of control sequences over the scenario tree, leading to a dual model predictive control formulation.
\end{abstract}


\end{frontmatter}

\section{Introduction}

Many advanced control methods require a model of the system to be controlled. The standard procedure is to perform offline system identification~(\cite{LL86}) of the plant and develop a posteriori a controller based on the identified model. This approach, however, assumes that all system characteristics have been observed during offline identification, which is becoming increasingly challenging and time-consuming for complex systems and environments. Adaptive control techniques~(\cite{KJA94}) address this problem by simultaneously learning and controlling an uncertain dynamical system. This leads to the well-known exploration-exploitation trade-off, which is computationally difficult to address. For this reason, many techniques revert to passive learning, i.e. the control inputs are chosen only with respect to the performance objective and not in order to drive active exploration of the unknown dynamics.

In this paper, we develop an approximate dual control
algorithm that aims at optimally balancing active exploration and exploitation in the optimal control of partially unknown systems with continuous state and input spaces, while offering a tractable optimization-based formulation. 
The proposed method casts the exploration-exploitation trade-off as a
stochastic optimal control problem by extending the state with the
current parameter estimate, for which the optimal solution can be
determined by applying stochastic dynamic programming (DP) with imperfect
state information~(\cite{DB17}). 
The solution relies on the information state of the
system, the dimensionality of which grows rapidly along the
horizon. For discrete state, input and parameter spaces this results in partially observable Markov decision processes (POMDP)~(\cite{PP06}), which are hard to solve even for small and medium sized systems~(\cite{JCS97}).

Stochastic DP provides the optimal exploration-exploitation trade-off with respect to the specified cost, and can therefore be regarded as the exact solution to the dual control problem~(\cite{AAF60}, \cite{AM17}). However, exact DP solutions are computationally tractable only for discrete state and action spaces of very small dimension. Among the approximate solutions proposed in the literature, dual control methods based on approximations of DP are known as \emph{implicit} dual control methods~(\cite{NMF00}), since
the dual effect is implicit in the approximate solution of the Bellman equation.
Solutions can be found by using approximate dynamic programming (ADP)~(\cite{ET73}, \cite{DB85}) or the wide-sense property~(\cite{EK16}). An alternative is given by
\emph{explicit} dual control methods, which provide the dual effect
explicitly, e.g. by heuristically adding probing features to the control
inputs or to the cost~(\cite{GM14}, \cite{MT2014}, \cite{TANH17}). We refer to~(\cite{AM17}) for more references and an overview of the literature.

The formulation proposed in this paper is an implicit dual control technique that makes use of an ADP strategy, generally referred to as a rollout approach~(\cite{DB17}). The method is based on splitting the control
horizon in a dual and an exploitation part. In the dual part, we
approximate the DP solution by constructing a
scenario tree from samples of the uncertain system parameters and process
noise. Each subtree is associated with a control input and a parameter distribution conditioned on the previously seen realizations and obtained at each time step by Bayesian estimation. A crucial feature of this formulation is its ability to
maintain the dual effect by continuously updating the
parameter distribution in the prediction based on scenarios. 
In the exploitation part, the parameter distribution is
fixed for the remaining time steps of the control horizon. Furthermore, the control sequence
associated with each sampled trajectory is optimized in open-loop, as in model predictive control (MPC). Similar techniques developed in an MPC setting are presented in~(\cite{ST18}), in which future observations are used to update disturbance confidence bounds in a robust formulation, and~(\cite{KGH15}), in which the unknown system parameter samples are updated using an ensemble Kalman filter. 

In contrast to these approaches, we systematically derive a dual control method with regard to the underlying stochastic optimal control problem and its exact DP solution.  Sample-based computations approximate the expected value operations in the Bellman equation, resulting in the construction of a scenario tree. Within each scenario branch, we update the available information by Gaussian conditioning of the parameter distribution on observations. This explicit connection to DP and Bayesian estimation comes with several advantages. It allows the connection to a number of established ADP strategies, for instance motivating the use of a rollout approach, which provides computational tractability for longer control horizons. This interpretation also provides a statistically consistent analysis of concepts introduced in~(\cite{KGH15}). Furthermore, a Bayesian update of the parameter distribution within each scenario branch provides a flexible framework. In this paper, we specifically address parameter affine systems under Gaussian noise, for which the Bayesian update can be analytically expressed. By providing a clear interpretation of different approximation steps in the DP framework, the proposed method can be easily extended to other system classes and distributions. Finally, we show that the overall optimization problem, involving both the dual and exploitation part, can be formulated with respect to a collection of control sequences along the branches of the scenario tree, resulting in a dual stochastic MPC approach which can be solved using gradient-based optimization techniques.

The paper is organized as follows. In Section~\ref{preliminaries} we formulate the problem and recount the stochastic dynamic programming solution. The proposed approximate dual control method is presented in Section~\ref{sec_DRSMPC}. In Section~\ref{simulations} we discuss examples and in Section~\ref{conlusions} final remarks.

\subsection{Notation}
With $\text{P}[\cdot|\cdot]$ we refer to the conditional probability. A normally distributed variable $x$ with mean $\mu$ and covariance $\Sigma$ is denoted $x \sim \mathcal{N}(\mu,\Sigma)$. $\mathbb{E}_x[\cdot]$ represents the expected value with respect to the random variable $x$. We use upper script indices to refer to samples of a quantity, and lower script indices for the time step, i.e. $x_k^{j}$ denotes a sample $j$ of state variable $x$ at stage $k$. The zero matrix is defined as $0_{n,m}$ and the identity matrix as $\mathbb{I}_{n,m}$, with $n$ rows and $m$ columns.

\section{Preliminaries}
\label{preliminaries}

\subsection{Problem Formulation}

We consider the control of discrete-time dynamical systems with parametric uncertainty subject to additive process noise, which can be described by
\begin{equation}
x_{k+1} = \Phi(x_k,u_k) \theta + w_k,
\label{eq:system}
\end{equation}
where $\Phi(x_k,u_k) \in \mathbb R^{n_x \times n_b}$ is a matrix composed of
nonlinear basis functions $\phi_{ij}(x_k,u_k)$  mapping the states $x_k \in
\mathbb{R}^{n_x}$ and control inputs  $u_k \in \mathbb{U} \subseteq \mathbb{R}^{n_u}$ to scalar values,
i.e. $\phi_{ij}:\mathbb R^{n_x + n_u} \rightarrow \mathbb R$. The set $\mathbb{U}$ defines the input constraints, the disturbance
$w_k \sim \mathcal{N}(0,\Sigma_w)$ is assumed to be i.i.d. Gaussian and $\theta
\in \mathbb R^{n_b}$ is a vector of fixed but uncertain parameters. In this
paper, we consider the case of Gaussian distributed continuous parameters
$\theta \sim \mathcal{N}(\mu_{\theta},\Sigma_{\theta})$. 
 Note that system~\eqref{eq:system} includes, e.g., the case of LTI systems with parametric uncertainty.

The goal is to find an optimal policy sequence $\Pi = \{ \pi_0, \dots, \pi_{N-1}
\}$ for system~\eqref{eq:system}, where $\pi_k : \mathbb{R}^{n_x} \rightarrow
\mathbb{U}$, minimizing the finite horizon cost
\begin{equation}
J_N(\Pi,x_0) := \mathbb{E}_{\theta,w_0,\dots,w_{N-1}} \Bigg [ \sum_{k=0}^{N-1} l_k(x_k,\pi_k(x_k)) + l_N(x_N)\Bigg ],
\label{eq:cost_function}
\end{equation}
where $N$ is the length of the control horizon, $l_k:\mathbb{R}^{n_x} \times
\mathbb{R}^{n_u} \rightarrow \mathbb{R}$ is a potentially time-varying stage
cost and $l_N:\mathbb{R}^{n_x}  \rightarrow \mathbb{R}$ is the terminal cost.

We address this finite-horizon stochastic optimal control problem using stochastic dynamic programming (DP), outlined in the following, offering the key property of inherently leading to a dual control policy. 

\subsection{Stochastic DP with Parametric Uncertainty}

In order to apply DP to problem~\eqref{eq:cost_function}, we describe system~\eqref{eq:system} using the augmented state $\begin{bmatrix}
x_k,  \theta_k
\end{bmatrix}^T$, i.e.
\begin{equation}
\begin{bmatrix}
x_{k+1} \\ \theta_{k+1}
\end{bmatrix}  = \begin{bmatrix}
 \Phi(x_k,u_k)\theta_{k} \\ \theta_{k} 
\end{bmatrix} + 
\begin{bmatrix}
w_k \\ 0_{n_b,1}
\end{bmatrix} ,
\label{eq:aug_system}
\end{equation}
for which information regarding the constant and unknown parameter state $\theta_{k}=\theta$ is only accessible through measurements of the state $x_k$. This formulation allows for phrasing the problem in the framework of systems with imperfect state information~(\cite{DB17}), for which the DP solution should take into account the information about the parameter distribution gained along the control horizon. The information available at time step $k$ can be formalized using the information vector $I_k$, defined as
\begin{equation*}
I_k := [ x_k,\dots,x_0,u_{k-1},\dots,u_0 ],
\end{equation*}
with $I_0 := [ x_0 ]$. DP recursively computes the optimal policy by evaluating the Bellman equation (see e.g.~\cite{RB66}) as a function of the information vector $I_k$, i.e.
\begin{equation}
J^*_k(I_k) :=  \displaystyle\min_{\pi_k} \; l_k(x_k, \pi_k) + \mathbb{E}_{\theta,w_k} \Big[ J^*_{k+1}(x_{k+1}) \Big | I_k  \Big],
\label{eq:DP}
\end{equation}
for $k = 0, \dots, N-1$, which is initialized with $J^*_N(I_N) := l_N(x_N)$. The
expected value of the optimal cost-to-go is evaluated with respect to the
process noise and the unknown parameters, given the available information. The
evolution of the probability distribution of the parameter vector, conditioned
on the information state, can be recursively obtained via Bayesian 
estimation~(\cite{AM17})
\begin{equation}
 \text{P}[\theta|I_k] = \frac{\text{P}[x_k|\theta,u_{k-1},I_{k-1}]\text{P}[\theta|I_{k-1}]}{\text{P}[x_k|u_{k-1},I_{k-1}]} ,
\label{eq:Bayesian estimation}
\end{equation}
with $\text{P}[\theta | I_{0}] := \text{P}[\theta]$, i.e. the prior distribution over the parameter vector $\theta$. 

The resulting DP policy therefore takes into account future observations, which
affect the knowledge about the parameter distribution. As a consequence, the solution inherently explores the system as necessary to optimally
solve the control problem given by~\eqref{eq:cost_function}, providing
\emph{dual control} with optimal exploration-exploitation trade-off.
%
Unfortunately, the alternation of minimization and expectation steps induced by
the Bellman equation is generally computationally intractable even under the
given assumption that the unknown parameter is Gaussian distributed and the
state is perfectly measured~(\cite{EK16}), in which case~\eqref{eq:Bayesian
estimation} is analytically tractable.

In the next section, we propose an approximate dual control method based on stochastic DP, which provides a tractable formulation by splitting the control horizon in a dual part and exploitation part and approximating the solution via sampling.  

\section{Dual Stochastic MPC} 
 \label{sec_DRSMPC}

The approach introduced in the following relies on a suboptimal solution to the Bellman equation, often referred to as the rollout approach. Optimally solving problem~\eqref{eq:cost_function} requires to carry out the DP algorithm for all steps of the control horizon of length $N$. The idea of a rollout approach is to use a truncated horizon of length $L<N$ and approximating the cost to go $\tilde{J}_{L}(I_{L})$ for the remainder of the horizon. The approximate DP recursion is given by
\begin{equation}
\tilde J_k(I_k) :=  \displaystyle\min_{\pi_k} \; l_k(x_k, \pi_k) + \mathbb{E}_{\theta,w_k} \Big[ \tilde J_{k+1}(x_{k+1}) \Big | I_k  \Big] ,
\label{eq:approx_costfunction}
\end{equation}
for $k = 0, \dots, L-1$, which is initialized with $\tilde J_L(I_L)$ and results
in a $L$-step lookahead problem~(\cite{DB17}). The terminal cost
$\tilde{J}_{L}(I_{L})$ is evaluated with respect to a suboptimal base policy
$\tilde\pi_L$, resulting in a suboptimal control policy with
respect to the $N$-step horizon cost~\eqref{eq:cost_function}. It can, however,
be shown that the policy obtained from~\eqref{eq:approx_costfunction} is always
improving with respect to the base policy~(\cite{DB17}). For this reason, rollout
approaches are generally implemented in a receding horizon fashion. 

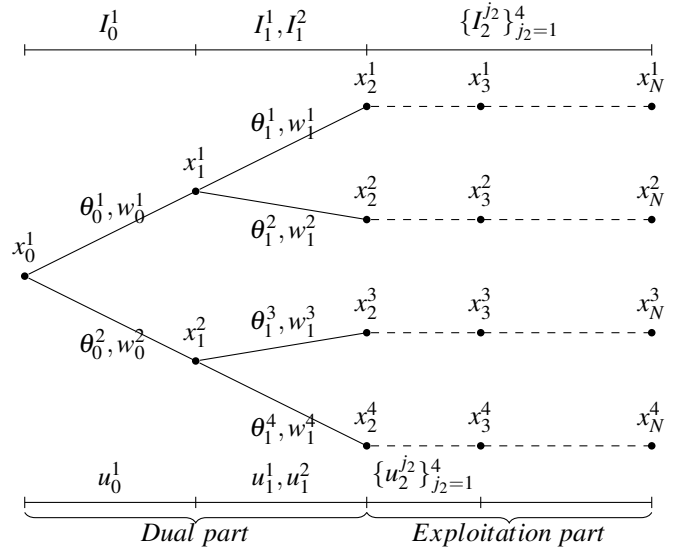
\begin{figure}
      \centering
\begin{tikzpicture}[c/.style 2 args={insert path={node[n={#1}{#2}] (n#1#2){}}},
n/.style 2 args={circle,fill,inner sep=1pt,label={90:$x_{#1}^{#2}$}}, scale={0.75}
]
\path (-2,2)[c={0}{1}];
\draw (n01) -- node[above] {$\theta_0^1,w_0^1$} (1,3.5)[c={1}{1}];
\draw (n01) -- node[below] {$\theta_0^2,w_0^2$}(1,0.5)[c={1}{2}];
\draw (n11) -- node[above] {$\theta_1^1,w_1^1$}(4,5)[c={2}{1}] ;
\draw (n11) -- node[below] {$\theta_1^2,w_1^2$}(4,3)[c={2}{2}] ;
\draw (n12) -- node[above] {$\theta_1^3,w_1^3$}(4,1)[c={2}{3}] ;
\draw (n12) -- node[below] {$\theta_1^4,w_1^4$}(4,-1)[c={2}{4}] ;
\draw[dashed] (n21) -- (6,5) [c={3}{1}] -- (9,5) [c={N}{1}];
\draw[dashed] (n22) -- (6,3) [c={3}{2}] -- (9,3) [c={N}{2}];
\draw[dashed] (n23) --(6,1) [c={3}{3}] -- (9,1) [c={N}{3}];
\draw[dashed] (n24)  -- (6,-1) [c={3}{4}] -- (9,-1) [c={N}{4}];
\draw (-2,-2) -- node[above] {$u_0^1$}(1,-2) -- node[above] {$u_1^1,u_1^2$}(4,-2) -- node[above] {$\{u_2 ^{j_2}\}_{j_2 = 1}^{4}$}(6,-2) -- (9,-2);    
   \foreach \y in {-2,1,4,6,9}{
     \draw (\y,-2.1) -- (\y,-1.9);
     }
\draw (-2,6) -- node[above] {$I_0^1$}(1,6) -- node[above] {$I_1^1,I_1^2$}(4,6) -- node[above] {$\{I_2^{j_2}\}_{j_2=1}^{4}$}(9,6);    
   \foreach \y in {-2,1,4,9}{
     \draw (\y,5.9) -- (\y,6.1);
     }
\draw [decorate,decoration={brace,amplitude=4pt,mirror},yshift = -5]
      (-2,-2) -- (4,-2) node [midway,yshift=-0.3cm] {$Dual \; part$};
\draw [decorate,decoration={brace,amplitude=4pt,mirror},yshift = -5]
      (4,-2) -- (9,-2) node [midway,yshift=-0.3cm] {$Exploitation \; part$};                               
\end{tikzpicture}
\caption{Scenario tree for $L=2$: The state propagation starts at a given initial condition and evolves, under a given input sequence, according to the $N_s=2$ sampled parameters and disturbances. The exploitation part further predicts the trajectory using the parameter distribution from step $k=2$.}
\label{fig:Figure1}
\end{figure}

We propose an approximate dual control algorithm for
solving~\eqref{eq:cost_function} based on the principles of the rollout
approach. 
The control horizon is split into a dual part of length $L$, which is
formulated as an $L$-step lookahead problem~\eqref{eq:approx_costfunction}, and
an exploitation part of length $N-L$, which is captured via the terminal
cost-to-go $\tilde J_L(I_L)$. While the rollout approximation
in~\eqref{eq:approx_costfunction} maintains the property that the solution is
inherently dual, the stochastic DP algorithm is still computationally intractable. We address
this problem by further approximating~\eqref{eq:approx_costfunction} for the
dual part and  evaluating the expectations as averages over samples of process
noise and unknown parameters. These samples are used to build a scenario
tree, as exemplified in Figure~\ref{fig:Figure1} for $L=2$. The subtree for
every time step is defined by a control input and an updated information state,
which in turn affects the parameter distribution and subsequently samples of the
parameters for $k=0,\dots,L-1$. 

Using averages instead of evaluating the expected value allows for unnesting the minimizations arising in~\eqref{eq:approx_costfunction} and for simultaneously optimizing all control inputs associated to each subtree, without having to explicitly carry out the DP recursion. Nevertheless, this approximation still provides a solution that depends on future observations, while reacting to the sampled state associated with each subtree, hence ensuring dual control.

The terminal cost $\tilde J_{L}(I_L)$ is obtained from a base policy, which fixes
the information state and optimizes over a single control sequence for each
sampled trajectory for the remainder of the horizon of length $N - L$,
corresponding to a (non-dual) stochastic MPC problem. Overall, this formulation of the dual and the exploitation part allows for merging the two optimization subproblems into one optimization problem that computes a collection of control sequences for the entire control horizon $N$,  for all the branches of the scenario tree.

In the following subsections, further details are given on the scenario tree for solving the dual part and on the parameter inference~\eqref{eq:Bayesian estimation} performed under the assumption of a Gaussian distributed parameter. Furthermore, we discuss the optimal control problem for the exploitation part defining the terminal cost-to-go $\tilde J_L (I_L)$ and finally the overall optimization problem that merges the two parts and approximates the dual control problem in the form of a dual MPC approach.

\subsection{Dual Part }

\subsubsection{Scenario Tree Generation.} 

In order to approximate the DP recursion in the dual part, we consider a scenario tree generated by repeated sampling of the noise $w_k$ and parameter
vector $\theta$, given the distribution defined by the available information
vector $I_k$. The nodes of the tree are denoted by the state samples $x_k^{j_k}$, where step $k$ indicates the depth level of the tree and $j_k = 1,\dots,N_s^{k}$ indicates a sample at this level, with $N_s^k$ being the $k$-th power of the number of scenarios $N_s$.
At each node $x_k^{j_k}$ in the tree, we generate $N_s$ scenarios
of $w_k \sim \mathcal{N}(0,\Sigma_w)$ and $\theta \sim \text{P}[\theta |
I_k^{j_k}]$, that is the distribution of $\theta$ according to the information
vector $I_k^{j_k}$ at that node. This generates the child nodes according to~\eqref{eq:system}, i.e.
\begin{equation}
x_{k+1}^{j_{k+1}} = \Phi(x_k^{j_k},u_k^{j_k}) \theta^{j_{k+1}}_k + w^{j_{k+1}}_k, \quad j_{k+1} = 1,\dots,N_s^{k+1} 
\label{eq:lookahead}
\end{equation}
where node $j_{k}$ at step $k$ is the parent of node $j_{k+1}$ at step $k+1$, i.e. $x_k^{j_k} = p(x_{k+1}^{j_{k+1}})$.
At the first time step, for instance, we draw $N_s$ sample pairs $(w_0^{j_1},\theta_0^{j_1})$ according
to $w_0 \sim \mathcal{N}(0,\Sigma_w)$ and the prior distribution $\theta \sim
\text{P}[\theta]$. Starting from $x_0^1$ and applying an input $u_0^1$, this
gives rise to $N_s$ child nodes $x_1^{j_1}$, $j_1 = 1,\dots,N_s$, with information $I_1^{j_1} =
[x_0^1, u_0^1, x_1^{j_1}]$. Applying this procedure over $L$ steps, we obtain the scenario tree depicted in Figure \ref{fig:Figure1} for $L=2$ and $N_s = 2$.

The updated distributions $\text{P}[\theta | I_k^{j_k}]$ are computed for each child node using Bayes rule~\eqref{eq:Bayesian estimation}, enabling recursive application of the procedure, which will be detailed in the following. The scenario tree depends on the selection of inputs $u_k^{j_k}$ which, as opposed to a classical DP recursion, will be optimized in one batch (see subsection
\ref{DP_scenario} for further details).

\subsubsection{Parameter Update.} 
\label{parameter update}

The vector of available information at each node of the scenario tree is given by 
\begin{equation*}
I_{k+1}^{j_{k+1}} = [x_{k+1}^{j_{k+1}},u_{k}^{j_{k}},I_{k}^{j_{k}}], \quad \text{with }I_1^0 = x_0^1
\end{equation*}
such that for the element $x_k^{j_k}$ of $I_k^{j_k}$, the relation $x_k^{j_k} =
p(x_{k+1}^{j_{k+1}})$ holds. This vector collects the states visited up to time $k$ and inputs applied until $k-1$ for each node $j_k$, and is used to update the probability distribution of the unknown parameters $\theta$ using Bayesian estimation~\eqref{eq:Bayesian estimation}. The case considered here,
where in~\eqref{eq:system} the uncertain parameters affect the system linearly
and the noise as well as the prior parameter distribution is Gaussian, allows inference to be carried out in closed form.

Assuming that, given the current information vector $I_k^{j_k}$ at stage $k$ and
node $j_k$, the parameter vector is Gaussian distributed, i.e.
\[ 
  \text{P}[\theta | I_k^{j_k}] =
\mathcal{N}(\mu_{\theta_k}^{j_k},\Sigma_{\theta_k}^{j_k}), \]
we generate samples $\theta_k^{j_{k+1}}$ from $\text{P}[\theta | I_k^{j_k}]$ 
and find the likelihood of the state at the next time step as 
\begin{equation*}
\text{P}[x_{k+1}| \theta_k^{j_{k+1}},u_k^{j_k},I_k^{j_k}] = \mathcal{N}( \Phi(x_k^{j_k},u_k^{j_k})\theta_k^{j_{k+1}} ,\Sigma_w) \, .
\end{equation*}
By additionally drawing $N_s$ samples from the noise distribution and choosing an input
$u_k^{j_k}$, we generate realizations of the state $x^{j_{k+1}}_{k+1}$ at nodes
$j_{k+1}$, which define the information state $I^{j_{k+1}}_{k+1}$. According to Bayes'
rule~\eqref{eq:Bayesian estimation}, at these nodes we have a parameter distribution
\begin{equation}
\text{P}[\theta | I_{k+1}^{j_{k+1}}] 
\varpropto \text{P}[x_{k+1}| \theta_k^{j_{k+1}},u_k^{j_k},I_k^{j_k}] \; \text{P}[\theta | I_k^{j_k}].
\label{eq:update}
\end{equation}
Since the Gaussian distribution is self-conjugate, the posterior will also be Gaussian, with updated mean and covariance matrix defined as~(\cite{CB06})
\begin{equation}
\begin{split}
& \big [\Sigma_{\theta_{k+1}}^{j_{k+1}} \big ]^{-1} = \big [ \Sigma_{\theta_{k}}^{j_k} \big] ^{-1} + \Phi(x_k^{j_k},u_k^{j_k})^T \Sigma_w^{-1} \Phi(x_k^{j_k},u_k^{j_k}),  \\
& \mu_{\theta_{k+1}}^{j_{k+1}} = \Sigma_{\theta_{k+1}}^{j_{k+1}}(\big [ \Sigma_{\theta_{k}}^{j_k} \big ]^{-1} \mu_{\theta_{k}}^{j_k} + \Phi(x_k^{j_k},u_k^{j_k})^T \Sigma_w^{-1} x_{k+1}^{j_{k+1}} ). 
\end{split}
\label{eq:gaussian_update}
\end{equation}
This provides a recursive update of the first and second moments of the parameter
distribution, based on the information available at the last measured node $I_{k+1}^{j_{k+1}}$. Starting this procedure from a Gaussian prior
$\text{P}(\theta)$ therefore provides an analytic recursion, resulting in Gaussian parameter estimates at each node in the
scenario tree.

This Bayesian framework provides great flexibility and can be readily adjusted to different use cases, for instance when the updated distributions cannot be analytically computed. Examples found in~(\cite{SS15}, \cite{KGH15}) make use of unscented and ensemble Kalman filters, which can be understood as an approximate Bayesian parameter update.

\subsubsection{DP Approximation using Scenario Tree.} 
\label{DP_scenario}
We outline the procedure for solving~\eqref{eq:approx_costfunction} using the scenario tree and assuming that $\tilde J_L(I_L^{j_L})$ is given for each sampled path. The formulation of the terminal cost-to-go will then be discussed in subsection~\ref{exploitation}. The expectation in~\eqref{eq:approx_costfunction} is approximated at each step as an average sum over realizations of process noise and unknown parameters, drawn from the updated distribution~\eqref{eq:update}, i.e.
\begin{equation}
\begin{split}
& \tilde J_0(I_0^1)  := \displaystyle \min_{u_0^1} \; l_0(x_0^1,u_0^1) + \frac{1}{N_s}\sum_{j_1=1}^{N_s} \big [ \displaystyle \min_{u_1^{j_1}} \; l_1(x_1^{j_1},u_1^{j_1})   \\
& + \frac{1}{N_s^2}\sum_{j_2=1}^{N_s^2} \big [ \dots  + \displaystyle\frac{1}{N_s^{L-1}}\sum_{j_{L-1}=1}^{N_s^{L-1}}[\min_{u_{L-1}^{j_{L-1}}} \; l_{L-1}(x_{L-1}^{j_{L-1}},u_{L-1}^{j_{L-1}})  \\
& + \frac{1}{N_s^L}\sum_{j_L=1}^{N_s^L} \tilde J_L(I_L^{j_L})] \dots ]].
\end{split}
\label{eq:pre_rolling_horizon}
\end{equation}
Differently from problem~\eqref{eq:approx_costfunction}, which requires the evaluation of nested minimizations and expectations for $L$ steps, the sampled expected value allows to unnest the minimizations
\begin{equation}
\begin{split}
&\tilde J_0(I_0^1) := \displaystyle \min_{\bold{u}_0,\dots,\bold{u}_{L-1}}  l_0(x_0^1,u_0^1) + \frac{1}{N_s}\sum_{j_1=1}^{N_s} l_1(x_1^{j_1},u_1^{j_1}) + \\
& + \frac{1}{N_s^2}\sum_{j_2=1}^{N_s^2} l_2(x_2^{j_2},u_2^{j_2}) + \dots + \frac{1}{N_s^L}\sum_{j_L=1}^{N_s^L} \tilde J_L(I_L^{j_L}),
\end{split}
\label{eq:rolling_horizon}
\end{equation}
where $\bold{u}_k = \{ u_k^j \}_{j=1}^{N_s^k}$ is the collection of control inputs at time $k$ for all sampled trajectories $N_s^k$. Note that this reformulation is possible since cost functions in outer minimizations are independent of the  inner minimizations. This can be easily seen by considering the simple example of minimizing functions $f_1(\cdot), f_2(\cdot)$ with respect to variables $x_1,x_2$:
$$\displaystyle\min_{x_1} f(x_1) + \min_{x_2} f(x_1,x_2) = \displaystyle\min_{x_1,x_2} f(x_1) + f(x_1,x_2).$$
The reformulation in~\eqref{eq:rolling_horizon} avoids the explicit DP recursion and enables simultaneous minimization of the control sequences associated with each sampled path. 


\subsection{Exploitation Part}
\label{exploitation}

In the exploitation part, we fix the distribution of the parameters in the
prediction and solve a simplified problem minimizing the expected cost for the remainder of the horizon over an input sequence $u_{L:N-1}^{j_L} =
\{u_L^{j_L},\dots,u_{N-1}^{j_L} \}$, giving rise to the terminal cost
$\tilde J_L(I_L^{j_L})$ in problem \eqref{eq:rolling_horizon}:
\begin{align}
  \tilde J_L(I_L^{j_L}) &= \min_{u_{L:N-1}^{j_L}} J_L(I_L^{j_L},u_{L:N-1}^{j_L}) \, , \label{eq:tail_cost}\\
J_L(I_L^{j_L},u_{L:N-1}^{j_L}) &= \displaystyle\mathbb{E}_{\theta,w_L,\dots,w_{N-1}}\! \Bigg[ \! \sum_{k=L}^{N-1} l_k(x_k^{j_L},u_k^{j_L}) + l_N(x_N^{j_L}) \Bigg |  I_L^{j_L} \! \Bigg] \, . \nonumber
\end{align}
The cost-to-go is thereby implicitly defined and can be directly integrated with the optimization-based formulation of the approximate dual problem, as summarized in \ref{Approx_dual_control_problem}.

In the general case, the expected value needs to be numerically approximated,
e.g. again by sampling. For some cost functions, for instance the commonly used
quadratic cost, analytical expressions of the expected value are available in
terms of mean and variance information. Other possible choices with this
property are, e.g., linear or saturating cost functions~(\cite{MPD11}). 
We focus here on the quadratic case, for which the expected value can be
computed as 
\begin{equation*}
\begin{split}
\mathbb{E}_{x_k}[\; l(x_k,u_k)\;] &= \mathbb{E}_{x_k} [ \;||x_k||^2_Q \; ]+ ||u_k||^2_R \\ 
&= ||\mu_{x_k}||^2_Q + tr(Q\Sigma_{x_k}) + ||u_k||^2_R .
\end{split}
\end{equation*}
Even under the assumption of a Gaussian distributed parameter vector $\theta$, the
state will not remain normally distributed due to the product with the basis
function matrix, such that mean and variance
need to be approximated. We consider two common approaches: a certainty equivalent (CE) approach  and a first order Taylor approximation of~\eqref{eq:aug_system} around the mean value of the state. In the rest of this subsection we refer to one path index and omit it for the sake of simplicity.

For the certainty equivalent approach, consider system~\eqref{eq:aug_system} evaluated at the mean values, i.e.
\begin{equation}
\begin{bmatrix}
\mu_{x_{k+1}} \\ \mu_{\theta_{k+1}}
\end{bmatrix}  = \begin{bmatrix}
 \Phi(\mu_{x_k},u_k)\mu_{\theta_{L}} \\ \mu_{\theta_{L}} 
\end{bmatrix}, 
\label{0th-Taylor}
\end{equation}
where $ \mu_{\theta_{L}}$ is the mean obtained in the last step of the dual part~\eqref{eq:update}.  For the first-order Taylor approximation, the state dynamics are given by 
\begin{equation*}
\begin{split}
\begin{bmatrix}
x_{k+1} \\ \theta_{k+1}
\end{bmatrix} & = \begin{bmatrix}
 \Phi(\mu_{x_k},u_k)\mu_{\theta_{L}}\\ \mu_{\theta_{L}}
\end{bmatrix} + 
\begin{bmatrix}
w_k \\ 0_{n_b,1}
\end{bmatrix}  \\
& + \begin{bmatrix}
\nabla_x [\Phi(\mu_{x_k} , u_k)  \mu_{\theta_L}] & \Phi(\mu_{x_k} , u_k) \\ 0_{n_b,n_x} & \mathbb{I}_{n_b,n_b}
\end{bmatrix}
\begin{bmatrix}
x_k - \mu_{x_k} \\ \theta_k - \mu_{\theta_L}
\end{bmatrix},
\end{split}
\end{equation*}
providing simple update equations for both mean and covariance based on properties of affine transformations of Gaussian distributed variables:
\begin{equation}
\begin{split}
 \begin{bmatrix}
\mu_{x_{k+1}} \\ \mu_{\theta_{k+1}}
\end{bmatrix}  &= \begin{bmatrix}
 \Phi(\mu_{x_k},u_k)\mu_{\theta_{L}} \\ \mu_{\theta_{L}} 
\end{bmatrix}  \\
 \Sigma_{k+1} &= \bar \Sigma_w + \bar A \Sigma_k \bar A^T,
\end{split}
\label{1st-Taylor}
\end{equation}
where 
\begin{equation*}
\begin{split}
\Sigma_k & = \begin{bmatrix}
\Sigma_{x_k} & \Sigma_{x_k,\theta_k} \\
\Sigma_{x_k,\theta_k}^T & \Sigma_{\theta_k} 
\end{bmatrix} \\
\bar A & = \begin{bmatrix}
\nabla_x [\Phi(\mu_{x_k} , u_k)  \mu_{\theta_L}] & \Phi(\mu_{x_k} , u_k) \\ 0_{n_b,n_x} & \mathbb{I}_{n_b,n_b}
\end{bmatrix} \\
\bar \Sigma_w & = \begin{bmatrix}
\Sigma_w & 0_{n_x,n_b} \\
0_{n_b,n_x} & 0_{n_b,n_b}
\end{bmatrix}.
\end{split}
\end{equation*}
Therefore, the prediction in \eqref{eq:tail_cost} can be carried out using either \eqref{0th-Taylor} or \eqref{1st-Taylor} for steps $k = L, \dots, N-1$, with initialization at step $L$ being 
\begin{equation*}
\begin{split}
 \begin{bmatrix}
\mu_{x_{L}} \\ \mu_{\theta_{L}}
\end{bmatrix}  &= \begin{bmatrix}
 x_L \\ \mu_{\theta_{L}} 
\end{bmatrix}  \\
 \Sigma_{L} &= \begin{bmatrix}
0_{n_x,n_x} & 0_{n_x,n_b} \\
0_{n_b,n_x} & \Sigma_{\theta_L} 
\end{bmatrix},
\end{split}
\end{equation*}
where $\Sigma_{\theta_L}$ is the covariance obtained in the last step of the dual part~\eqref{eq:update}.
 
Based on the presented approximations of the expected value in the cost-to-go~\eqref{eq:tail_cost}, we can formulate the overall optimization problem in the next subsection.

\subsection{Final Approximate Dual MPC Problem}
\label{Approx_dual_control_problem}

The formulation of the dual and exploitation part in the form of an optimization problem allows for merging the two subproblems into one optimization problem with respect to a collection of input sequences along the control horizon of length $N$:
\vspace{-0.5em}
\begin{mini}[2]{\bold{u}_0,\dots,\bold{u}_{N-1}}{ \! \! \! \! \sum_{k=0}^{L-1} \frac{1}{N_s^k}\sum_{j_k=1}^{N_s^k} l_k(x_k^{j_k},u_k^{j_k}) + \frac{1}{N_s^L}\sum_{j_L=1}^{N_s^L} J_L(I_L^{j_L},u_{L:N-1}^{j_L})}{\label{eq:optimization}}{}
\addConstraint{\! \! \! \! \eqref{eq:lookahead}}{}{\quad k=0,\dots,L-1}
\addConstraint{\! \! \! \! \theta_k^{j_{k+1}} \text{ drawn from \eqref{eq:update}} }{}{\quad k=0,\dots,L-1}
\addConstraint{\! \! \! \!  \eqref{0th-Taylor} \text{ or } \eqref{1st-Taylor} }{}{\quad k=L, \dots, N-1}
\addConstraint{\! \! \! \! \bold{u}_k \in \mathbb{U}^k }{}{\quad k=0, \dots, N-1}
\end{mini}
Problem~\eqref{eq:optimization} is solved in a receding horizon fashion, updating at each time step the distribution of the  parameters.
\begin{remark}
Note that input constraints are directly incorporated in \eqref{eq:optimization}. State constraints can similarly be added, however no closed-loop feasibility  guarantees are provided.
\end{remark}

It is important to note that in problem~\eqref{eq:optimization},  the parameter distribution~\eqref{eq:update} from which the samples are drawn at each stage, is a function of the optimization variables $\bold{u}_k$. By assuming the parameters to be Gaussian, we can however perform online sampling by generating at first $N_s^{k+1}$ offline realizations from a standard normal distribution $\mathcal{N}(0,\mathbb{I}_{n_b,n_b})$, and then linearly transforming them online with the analytical expressions of mean and variance~\eqref{eq:gaussian_update} for each parent node $j_k$, i.e.
\begin{equation*}
\theta_k^{j_{k+1}} = \mu_k^{j_k} + \text{chol}(\Sigma_k^{j_k})\bar \theta_k^{j_{k+1}},
\end{equation*}
where $\bar \theta_k^{j_{k+1}}$ is drawn from $\mathcal{N}(0,\mathbb{I}_{n_b,n_b})$, with chol($\cdot$) standing for the Cholesky decomposition. As a consequence, we can avoid the explicit resampling of the unknown parameter in problem \eqref{eq:optimization}, yet generating realizations from an updated distribution computed exactly via Bayesian estimation.

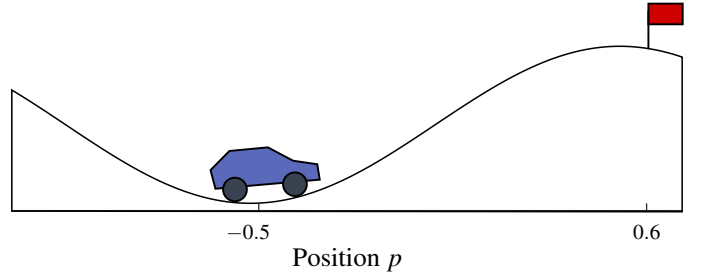
\begin{figure}[t]
\def\svgwidth{\linewidth}
\begingroup%
  \makeatletter%
  \providecommand\color[2][]{%
    \errmessage{(Inkscape) Color is used for the text in Inkscape, but the package 'color.sty' is not loaded}%
    \renewcommand\color[2][]{}%
  }%
  \providecommand\transparent[1]{%
    \errmessage{(Inkscape) Transparency is used (non-zero) for the text in Inkscape, but the package 'transparent.sty' is not loaded}%
    \renewcommand\transparent[1]{}%
  }%
  \providecommand\rotatebox[2]{#2}%
  \newcommand*\fsize{\dimexpr\f@size pt\relax}%
  \newcommand*\lineheight[1]{\fontsize{\fsize}{#1\fsize}\selectfont}%
  \ifx\svgwidth\undefined%
    \setlength{\unitlength}{335.74499161bp}%
    \ifx\svgscale\undefined%
      \relax%
    \else%
      \setlength{\unitlength}{\unitlength * \real{\svgscale}}%
    \fi%
  \else%
    \setlength{\unitlength}{\svgwidth}%
  \fi%
  \global\let\svgwidth\undefined%
  \global\let\svgscale\undefined%
  \makeatother%
  \begin{picture}(1,0.35)%
    \lineheight{1}%
    \setlength\tabcolsep{0pt}%
    \put(0,0){\includegraphics[width=\unitlength, page=1]{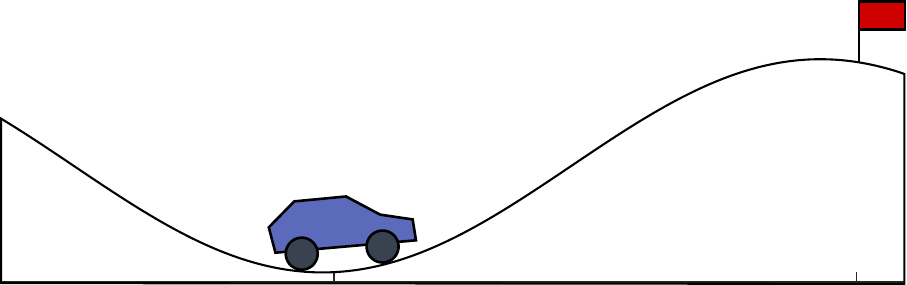}}%
    \put(0.32,-0.04){\color[rgb]{0,0,0}\makebox(0,0)[lt]{\lineheight{1}\smash{\begin{tabular}[t]{l}\small $-0.5$\end{tabular}}}}%
    \put(0.925,-0.04){\color[rgb]{0,0,0}\makebox(0,0)[lt]{\lineheight{1.25}\smash{\begin{tabular}[t]{l} \small $0.6$\end{tabular}}}}%
    \put(0.5,-0.08){\color[rgb]{0,0,0}\makebox(0,0)[ct]{\lineheight{1.25}\smash{\begin{tabular}[t]{l}  Position $p$\end{tabular}}}}%
  \end{picture}%
\endgroup%

\vspace{0.25cm}
\caption{The mountain car problem: A car is initially placed in a valley at position $-0.5$ and the goal is to pass a hill at position $0.6$. Depending on the parameter realization, the car is unable to climb the hill directly and needs to 'swing up'.}
\label{fg:mountainCar}
\end{figure}

\section{Simulation Examples}
\label{simulations}

We demonstrate the proposed algorithm with two different simulations. The first example is an illustrative scalar LTI system with unknown actuator gain and additive Gaussian process noise, for which we analyze the solution to problem~\eqref{eq:optimization} and provide closed-loop results. The second example is typically referred to as the mountain car problem~\citep{RSS98}, for which we compare the performance of our approach with respect to a certainty equivalent and an adaptive MPC controller. 

\subsection{Scalar example}

As a first illustrative example, we consider the control of a scalar system described by
\begin{equation*}
x_{k+1} = x_k + \theta u_k +w_k,
\end{equation*}
where $\theta$ is the unknown parameter, assumed to be distributed as $\theta \sim \mathcal{N}(1,10)$, $w_k \sim \mathcal{N}(0,0.1)$ is process noise, and the true parameter is $\theta_{\text{true}}=3$.
Linearity with respect to the unknown parameter allows inference to be carried out in closed-form~\eqref{eq:update}. Furthermore, for this simple set-up, propagation of state mean and variance in the exploitation part can be evaluated exactly, as a result of linearly combined Gaussian distributed variables. The goal is to regulate the system to the origin from an initial condition $x_0 = 5$. We define a quadratic cost function 
$l_k(x_k,u_k) = ||x_k||^2_{Q} + ||u_k||_{R}^2
$, where $Q = 10$ and $R = 0.01$, and solve~\eqref{eq:optimization} using $N_s = 50$ samples and $L=1$ lookahead steps. 

\begin{figure}[t]
  \centering
  	\hspace{-0.7em}\input{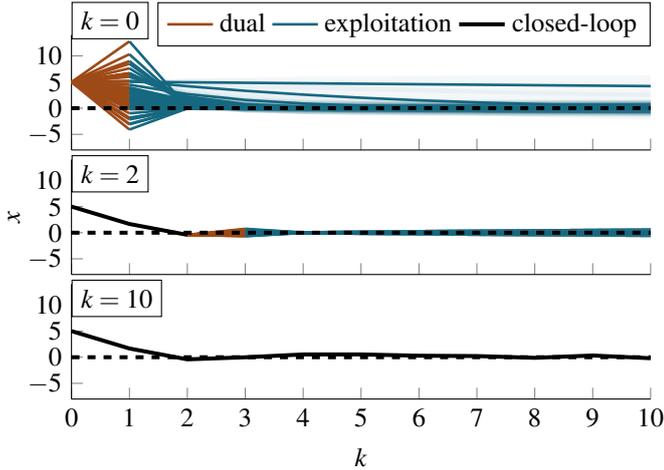}\vspace{-1em}
\caption{Regulation of LTI scalar system. Depicted are the closed-loop trajectory, mean and 1$\sigma$  bounds of the $N_s=50$ sampled open-loop trajectories. From the top: Plot 1 shows at time k=0 the predicted open-loop sequences. Plot 2 shows the closed-loop state for two time steps and the open-loop prediction generated at k=2. Plot 3 shows the closed-loop trajectory for 10 time steps. }
\label{fg:example1}
\end{figure}

The purpose of this illustrative example is to investigate the ability of the dual controller to actively explore and gain information about the unknown system parameter. Nevertheless, even this simple task would be very difficult for a non-dual controller, e.g. certainty equivalent MPC, since the direction of the actuator gain is unknown. Figure \ref{fg:example1} shows the predictions for both the dual and the exploitation part and the closed-loop solution at time steps $k=0,2,10$.  Notice that at time $k=0$, the control action is selected to test several strategies, including to go in the wrong direction with respect to the origin, in order to gain information about $\theta$. At time step $k=2$, the parameter is almost identified, and its updated distribution is $\theta \sim \mathcal{N}(3.1,0.07)$. The exploration is now reduced, and for the remaining time steps the controller regulates the system to the origin.

\subsection{Mountain Car}

As a second example we consider the mountain car problem with parametric model 
uncertainties. The goal is to drive a car from a valley past the top of a
hill, as illustrated in Figure~\ref{fg:mountainCar}. The dynamics of the system
are given by~\citep{RSS98}
\[
x_{k+1} = \begin{bmatrix} p_{k+1} \\ v_{k+1} \end{bmatrix} = \begin{bmatrix} p_{k} + T_s v_{k} \\ v_{k} - T_s \cos (3 p_{k}) \theta_1  + T_s u_k \theta_2  \end{bmatrix} + \begin{bmatrix} 0 \\ 1 \end{bmatrix} w_k \, ,
\]
where $p$ is the position and $v$ is the velocity of the vehicle, 
input $u$ is given by the acceleration and disturbances $w_k \sim \mathcal{N}(0, 0.001^2)$. We choose
a sampling time of $T_s =7$. 
The task in the mountain car problem is to climb a hill, characterized by the
target position $p_\text{target} = 0.6$ starting from an initial position $p_0 =
-0.5$ located in a valley. We express this goal by choosing a simple linear cost
function $l_k(x_k,u_k) = -p_k$, 
such that the system is encouraged to maximize its position as quickly as
possible, eventually exceeding the hill.

The parameters $\theta_1, \theta_2$ are
typically chosen such that the vehicle cannot reach the goal directly, but needs to 'swing back' in order to gain speed and then climb the
hill. We consider the case
where the parameters are uncertain, specifically 
\[ 
\begin{bmatrix} \theta_1 \\ \theta_2 \end{bmatrix} \sim
\mathcal{N}\left(\begin{bmatrix} 0.002 \\ 0.0025 \end{bmatrix},
\begin{bmatrix} 0.001^2 & 0 \\ 0 & 0.001^2 \end{bmatrix} \right) \, ,
\]
such that for some realizations swinging back is necessary, for others not.
In this setup, the optimal controller should first find the right strategy for the given
parameter realization and subsequently execute it to minimize the expected
cost.

To solve this task we generate an approximate dual controller, referred to as \emph{D-MPC}, with lookahead $L=3$
and $N_s = 5$ scenario realizations in each time step for an overall prediction
horizon of $N=15$.
We compare the \emph{D-MPC} to two variants, namely
\begin{description}
\item[\normalfont \emph{CE-MPC}] Certainty equivalent receding horizon
controller computing open-loop control sequences based on the prior maximum
likelihood parameter value $\mu_\theta$.
\item[\normalfont \emph{aS-MPC}] Adaptive stochastic receding horizon
controller. The controller uses $N_s^L$ parameter and noise
samples and computes the optimal input sequence by optimizing the sampled
average of the cost, approximating the expected value. The
controller passively learns and adapts the parameter knowledge in closed loop
using~\eqref{eq:gaussian_update}.  
\end{description}

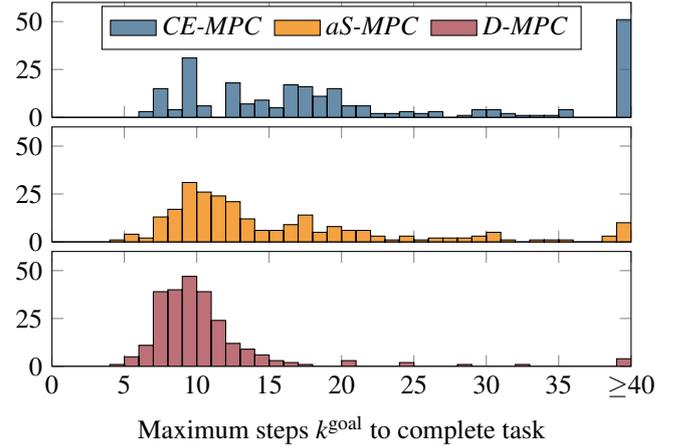
\begin{figure}[t]
  \centering
%
%
\definecolor{mycolor1}{rgb}{0.211764706,0.407843137,0.552941176}%
\definecolor{mycolor2}{rgb}{0.945098039,0.51372549,0.015686275}%
\definecolor{mycolor3}{rgb}{0.654901961,0.254901961,0.290196078}%
\begin{tikzpicture}

\begin{axis}[%
width=3in,
height=0.6in,
at={(0in,1.3in)},
scale only axis,
xmin=1,
xmax=41,
xtick = {1,6,11,16,21,26,31,36,41},
ymin=0,
ymax=60,
ytick = {0, 25, 50},
axis background/.style={fill=white},
xticklabels = {,,},
legend style={at={(0.5,0.97)},anchor=north},
legend columns=-1,
]
\addplot[ybar interval, fill=mycolor1, fill opacity=0.75, draw=black, area legend] table[row sep=crcr] {%
x	y\\
1	0\\
2	0\\
3	0\\
4	0\\
5	0\\
6	0\\
7	3\\
8	15\\
9	4\\
10	31\\
11	6\\
12	0\\
13	18\\
14	7\\
15	9\\
16	5\\
17	17\\
18	16\\
19	11\\
20	15\\
21	6\\
22	6\\
23	2\\
24	2\\
25	3\\
26	2\\
27	3\\
28	0\\
29	1\\
30	4\\
31	4\\
32	2\\
33	1\\
34	1\\
35	1\\
36	4\\
37	0\\
38	0\\
39	0\\
40  51\\
41  51\\
};
\addplot[ybar interval, fill=mycolor2, fill opacity=0.75, draw=black, area
legend] table[row sep=crcr] {%
x   y\\
0   0\\
};
\addplot[ybar interval, fill=mycolor3, fill opacity=0.75, draw=black, area
legend] table[row sep=crcr] {%
x   y\\
0   0\\
};
\addlegendentry{\emph{CE-MPC} \ }
\addlegendentry{\emph{aS-MPC} \ }
\addlegendentry{\emph{D-MPC}}
\end{axis}

\begin{axis}[%
width=3in,
height=0.6in,
at={(0in,0.65in)},
scale only axis,
xmin=1,
xmax=41,
xtick = {1,6,11,16,21,26,31,36,41},
ymin=0,
ymax=60,
ytick = {0, 25, 50},
axis background/.style={fill=white},
xticklabels = {,,},
legend style={legend cell align=left, align=left, draw=white!15!black}
]
\addplot[ybar interval, fill=mycolor2, fill opacity=0.75, draw=black, area legend] table[row sep=crcr] {%
x	y\\
1	0\\
2	0\\
3	0\\
4	0\\
5	1\\
6	4\\
7	2\\
8	13\\
9	17\\
10	31\\
11	26\\
12	24\\
13	21\\
14	12\\
15	6\\
16	6\\
17	9\\
18	14\\
19	5\\
20	8\\
21	6\\
22	6\\
23	3\\
24	1\\
25	3\\
26	1\\
27	2\\
28	2\\
29	2\\
30	3\\
31	5\\
32	1\\
33	0\\
34	1\\
35	1\\
36	1\\
37	0\\
38	0\\
39	3\\
40  10\\
41  10\\
};

\end{axis}

\begin{axis}[%
width=3in,
height=0.6in,
at={(0in,0in)},
scale only axis,
xmin=1,
xmax=41,
xtick = {1,6,11,16,21,26,31,36,41},
xticklabels = {0,5,10,15,20,25,30,35,$\geq$40 },
ymin=0,
ymax=60,
ytick = {0, 25, 50},
axis background/.style={fill=white},
legend pos = south east,
xlabel = {Maximum steps $k^{\text{goal}}$ to complete task},
legend style={legend cell align=left, align=left, draw=white!15!black}
]
\addplot[ybar interval, fill=mycolor3, fill opacity=0.75, draw=black, area legend] table[row sep=crcr] {%
x	y\\
1	0\\
2	0\\
3	0\\
4	0\\
5	1\\
6	5\\
7	11\\
8	39\\
9	40\\
10	47\\
11	39\\
12	24\\
13	12\\
14	9\\
15	6\\
16	3\\
17	2\\
18	1\\
19	0\\
20	0\\
21	3\\
22	0\\
23	0\\
24	0\\
25	2\\
26	0\\
27	0\\
28	0\\
29	1\\
30	0\\
31	0\\
32	0\\
33	1\\
34	0\\
35	0\\
36	0\\
37	0\\
38	0\\
39	0\\
40  4\\
41  4\\
};
\end{axis}

\end{tikzpicture}%
  \caption{Histogram of step number $k^{\text{goal}}$ to complete task in for the mountain car problem for different controller formulations, for 250 noise realizations. The finishing time $k^{\text{goal}}$ is the time at which the car position first exceeds $p_{\text{goal}} = 0.6$. }\label{fg:histogram}
\end{figure}

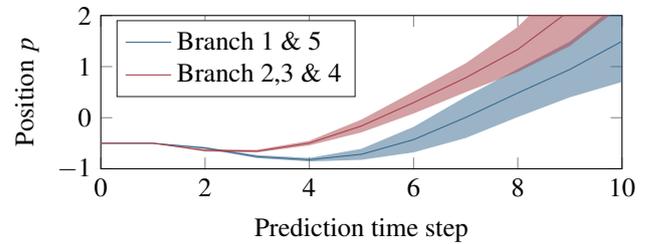
\begin{figure}
  \centering
%
%
\definecolor{mycolor1}{rgb}{0.211764706,0.407843137,0.552941176}%
\definecolor{mycolor3}{rgb}{0.654901961,0.254901961,0.290196078}%

\begin{tikzpicture}

\begin{axis}[%
width=2.7in,
height=0.8in,
scale only axis,
xmin=0,
xmax=10,
ymin=-1,
ymax=2,
xlabel = {Prediction time step},
ylabel near ticks,
ylabel = {Position $p$},
axis background/.style={fill=white},
legend style={legend cell align=left, align=left, draw=white!15!black},
legend pos = north west,
]
\addplot [color=mycolor1]
  table[row sep=crcr]{%
0	-0.5\\
1	-0.5\\
2	-0.59203950608305\\
3	-0.762486702747546\\
4	-0.822670469260442\\
5	-0.716526686351962\\
6	-0.429509280369436\\
7	0.00524324102816751\\
8	0.483378950217835\\
9	0.945353467110991\\
10	1.49486588929915\\
};
\addlegendentry{Branch 1 \& 5}

\addplot[area legend, draw=none, fill=mycolor1, fill opacity=0.5, forget plot]
table[row sep=crcr] {%
x	y\\
0	-0.5\\
1	-0.5\\
2	-0.572949984106229\\
3	-0.732057427447316\\
4	-0.787910592343254\\
5	-0.608238262871857\\
6	-0.179559259681575\\
7	0.408863819934402\\
8	0.953947132707329\\
9	1.49102195660801\\
10	2.29059412519061\\
10	0.699137653407696\\
9	0.399684977613973\\
8	0.0128107677283417\\
7	-0.398377337878067\\
6	-0.679459301057298\\
5	-0.824815109832067\\
4	-0.85743034617763\\
3	-0.792915978047777\\
2	-0.611129028059872\\
1	-0.5\\
0	-0.5\\
}--cycle;

\addplot [color=mycolor3]
  table[row sep=crcr]{%
0	-0.5\\
1	-0.5\\
2	-0.640183233641921\\
3	-0.654813630677979\\
4	-0.497146070140859\\
5	-0.162102542317961\\
6	0.295099841617649\\
7	0.78187524909506\\
8	1.33632867350138\\
9	2.10040680295903\\
10	3.05622737183892\\
};
\addlegendentry{Branch 2,3 \& 4}

\addplot[area legend, draw=none, fill=mycolor3, fill opacity=0.5]
table[row sep=crcr] {%
x	y\\
0	-0.5\\
1	-0.5\\
2	-0.617791530927189\\
3	-0.629101491129077\\
4	-0.452053526896765\\
5	-0.0364395054473786\\
6	0.509466131795328\\
7	1.06448001278144\\
8	1.77735601479651\\
9	2.79647344173833\\
10	3.92278058478349\\
10	2.18967415889435\\
9	1.40434016417973\\
8	0.895301332206241\\
7	0.499270485408682\\
6	0.0807335514399691\\
5	-0.287765579188544\\
4	-0.542238613384954\\
3	-0.680525770226881\\
2	-0.662574936356652\\
1	-0.5\\
0	-0.5\\
}--cycle;

\end{axis}

\end{tikzpicture}%
\caption{Prediction of the dual controller in the mountain car problem at
time step 0. Displayed are the mean and $2\sigma$ bounds of the $N_s^L=5^3$ planned trajectory samples, separated in two groups depending on the $5$ initial tree branches reflecting different strategies. The controller plans to swing back and depending on the collected information to start the hill climb at different points in time.}\label{fg:bifurcation}
\end{figure}

As performance indicator we investigate $k^{\text{goal}}$, i.e. the number of
time steps required until exceeding the target state $p_{\text{goal}} = 0.6$ for the
different controllers using $250$ simulation experiments with different noise
realizations. A histogram of the resulting distributions of $k^{\text{goal}}$ is
given in Figure~\ref{fg:histogram}. The results clearly show that the certainty
equivalent controller (\emph{CE-MPC}) is unable to exceed the hill within 40 time
steps and fails in a large portion of the trials. One can furthermore observe
two modes in the finishing times, the first around 10 time steps, corresponding to a success on first trial, and a second
starting around 15 time steps, corresponding to success after a first failure.
The adaptive stochastic controller (\emph{aS-MPC}) manages to complete most
trials within 40 time steps, since it is able to learn the parameter
distribution over time. The two modes, however, are still visible, meaning
that a number of climbing attempts by the \emph{aS-MPC} fail on first trial. The
dual controller (\emph{D-MPC}) in contrast is able to successfully climb the hill
on the first trial in almost all cases, exceeding the other controllers in performance.

Figure~\ref{fg:bifurcation} illustrates the prediction of the dual controller
in the first time step, which generates $N_s^L = 5^3$ sampled trajectories. By observing the related scenario tree, the $N_s=5$ branches can be grouped in
two distinct solution strategies, based on a short and longer swing-back. The
controller therefore plans to swing back, and depending on the information
gained, start the hill climb earlier or later. This flexibility in the planning, 
due to the ability of the controller to select a different control input
based on the information gained during execution, greatly helps to reduce conservatism.


\section{Conclusions}
We have presented an approximate dual control approach that is systematically derived from a stochastic DP framework and a rollout strategy. The formulation is based on separating the control horizon into a dual and an exploitation part. The dual part is formulated using a scenario tree with realizations of noise and unknown parameters, and the exploitation part optimizes over open-loop control sequences for the remainder of the horizon. By approximating the expected values using sampling, the two subproblems can be merged, simultaneously optimizing over a collection of input sequences along the control horizon. The problem of sampling online from a distribution that depends on the optimization variables is solved by generating samples from a standard normal distribution and affinely transforming them with the exact mean and covariance computed via Bayesian estimation over the prediction horizon. The proposed technique, thereby, offers a tractable procedure while maintaining the dual features of stochastic DP. The results were illustrated for a scalar LTI system and for the mountain car problem.
\label{conlusions}
%
%






%
%
%
%
%
%
%

\bibliography{bibliography}

\begin{thebibliography}{21}
\providecommand{\natexlab}[1]{#1}
\providecommand{\url}[1]{\texttt{#1}}
\providecommand{\urlprefix}{URL }
\expandafter\ifx\csname urlstyle\endcsname\relax
  \providecommand{\doi}[1]{doi:\discretionary{}{}{}#1}\else
  \providecommand{\doi}{doi:\discretionary{}{}{}\begingroup
  \urlstyle{rm}\Url}\fi

\bibitem[{{\AA}str{\"o}m and Wittenmark(2008)}]{KJA94}
{\AA}str{\"o}m, K.J. and Wittenmark, B. (2008).
\newblock \emph{Adaptive Control}.
\newblock Dover Publications, 2 edition.

\bibitem[{Bayard and Eslami(1985)}]{DB85}
Bayard, D.S. and Eslami, M. (1985).
\newblock Implicit dual control for general stochastic systems.
\newblock \emph{Optimal Control Applications and Methods}, 6(3), 265--279.

\bibitem[{Bellman(1966)}]{RB66}
Bellman, R. (1966).
\newblock Dynamic programming.
\newblock \emph{Science}, 153(3731), 34--37.

\bibitem[{Bertsekas(2017)}]{DB17}
Bertsekas, D.P. (2017).
\newblock \emph{Dynamic Programming and Optimal Control}.
\newblock Athena Scientific, 4th edition.

\bibitem[{Bishop(2006)}]{CB06}
Bishop, C.M. (2006).
\newblock \emph{Pattern Recognition and Machine Learning}.
\newblock Springer.

\bibitem[{Deisenroth and Rasmussen(2011)}]{MPD11}
Deisenroth, M.P. and Rasmussen, C.E. (2011).
\newblock {PILCO}: A model-based and data-efficient approach to policy search.
\newblock In \emph{28th Int. Conf. Machine Learning}, 465--472.

\bibitem[{Feldbaum(1961)}]{AAF60}
Feldbaum, A.A. (1961).
\newblock Dual control theory. {Part I} \& {II}.
\newblock \emph{Automation and Remote Control}, 21(9), 874--880, 1033--1039.

\bibitem[{Filatov and Unbehauen(2000)}]{NMF00}
Filatov, N. and Unbehauen, H. (2000).
\newblock Survey of adaptive dual control methods.
\newblock \emph{Control Theory and Applications}, 147, 118 -- 128.

\bibitem[{Hanssen and Foss(2015)}]{KGH15}
Hanssen, K.G. and Foss, B. (2015).
\newblock Scenario based implicit dual model predictive control.
\newblock \emph{Conf. Nonlinear Model Predictive Control}, 416 -- 421.

\bibitem[{Heirung et~al.(2017)Heirung, Ydstie, and Foss}]{TANH17}
Heirung, T.A.N., Ydstie, B.E., and Foss, B. (2017).
\newblock Dual adaptive model predictive control.
\newblock \emph{Automatica}, 80, 340--348.

\bibitem[{Klenske and Hennig(2016)}]{EK16}
Klenske, E.D. and Hennig, P. (2016).
\newblock Dual control for approximate {Bayesian} reinforcement learning.
\newblock \emph{J. Machine Learning Research}, 17(1), 4354--4383.

\bibitem[{Ljung(1986)}]{LL86}
Ljung, L. (1986).
\newblock \emph{System Identification: Theory for the User}.
\newblock Prentice-Hall.

\bibitem[{Marafioti et~al.(2014)Marafioti, Bitmead, and Hovd}]{GM14}
Marafioti, G., Bitmead, R.R., and Hovd, M. (2014).
\newblock Persistently exciting model predictive control.
\newblock \emph{Int. J. Adaptive Control and Signal Processing}, 28(6),
  536--552.

\bibitem[{Mesbah(2018)}]{AM17}
Mesbah, A. (2018).
\newblock Stochastic model predictive control with active uncertainty learning:
  A survey on dual control.
\newblock \emph{Annual Reviews in Control}, 45, 107 -- 117.

\bibitem[{Poupart et~al.(2006)Poupart, Vlassis, Hoey, and Regan}]{PP06}
Poupart, P., Vlassis, N., Hoey, J., and Regan, K. (2006).
\newblock An analytic solution to discrete {Bayesian} reinforcement learning.
\newblock In \emph{23rd Int. Conf. Machine Learning}, 697--704.

\bibitem[{Santamar\'{\i}a et~al.(1997)Santamar\'{\i}a, Sutton, and Ram}]{JCS97}
Santamar\'{\i}a, J.C., Sutton, R.S., and Ram, A. (1997).
\newblock Experiments with reinforcement learning in problems with continuous
  state and action spaces.
\newblock \emph{Adaptive Behavior}, 6(2), 163--217.

\bibitem[{Subramanian et~al.(2015)Subramanian, Lucia, and Engell}]{SS15}
Subramanian, S., Lucia, S., and Engell, S. (2015).
\newblock Economic multi-stage output feedback nmpc using the unscented kalman
  filter.
\newblock \emph{IFAC-PapersOnLine}, 48(8), 38 -- 43.
\newblock 9th IFAC Symposium on Advanced Control of Chemical Processes ADCHEM
  2015.

\bibitem[{Sutton and Barto(2018)}]{RSS98}
Sutton, R.S. and Barto, A.G. (2018).
\newblock \emph{Reinforcement Learning: An Introduction}.
\newblock MIT Press, 2 edition.

\bibitem[{Tanaskovic et~al.(2014)Tanaskovic, Fagiano, Smith, and
  Morari}]{MT2014}
Tanaskovic, M., Fagiano, L., Smith, R., and Morari, M. (2014).
\newblock Adaptive receding horizon control for constrained {MIMO} systems.
\newblock \emph{Automatica}, 50(12), 3019--3029.

\bibitem[{Thangavel et~al.(2018)Thangavel, Lucia, Paulen, and Engell}]{ST18}
Thangavel, S., Lucia, S., Paulen, R., and Engell, S. (2018).
\newblock Dual robust nonlinear model predictive control: A multi-stage
  approach.
\newblock \emph{J. Process Control}, 72, 39 -- 51.

\bibitem[{{Tse} and {Bar-Shalom}(1973)}]{ET73}
{Tse}, E. and {Bar-Shalom}, Y. (1973).
\newblock An actively adaptive control for linear systems with random
  parameters via the dual control approach.
\newblock \emph{Trans. Automatic Control}, 18(2), 109--117.

\end{thebibliography}

\end{document}